\documentclass[runningheads]{llncs}
\usepackage[T1]{fontenc}
\usepackage{multirow}
\usepackage{amsmath}
\usepackage{amssymb}
\usepackage{graphicx}
\begin{document}
\title{Could We Generate Cytology Images from Histopathology Images? An Empirical Study}
%
%
\author{Soumyajyoti Dey\inst{1,*} \and Sukanta Chakraborty \inst{2} \and
Utso Guha Roy\inst{2} \and
Nibaran Das\inst{1}}
\authorrunning{Dey et al.}
%
\institute{Jadavpur University, Kolkata, India \and
Theism Medical Diagnostics Centre, Kolkata, West Bengal, India
}
\maketitle              
\begin{abstract}

Automation in medical imaging is quite challenging due to the unavailability of annotated datasets and the scarcity of domain experts. In recent years, deep learning techniques have solved some complex medical imaging tasks like disease classification, important object localization, segmentation, etc. However, most of the task requires a large amount of annotated data for their successful implementation. To mitigate the shortage of data, different generative models are proposed for data augmentation purposes which can boost the classification performances. For this, different synthetic medical image data generation models are developed to increase the dataset. Unpaired image-to-image translation models here shift the source domain to the target domain. In the breast malignancy
identification domain, FNAC is one of the low-cost low-invasive modalities normally used by medical practitioners. But availability of public datasets in this domain is very poor. Whereas, for automation of cytology images, we need a large amount of annotated data.  Therefore synthetic cytology images are generated by translating breast histopathology samples which are publicly available. In this study, we have explored traditional image-to-image transfer models like CycleGAN, and Neural Style Transfer. Further, it is observed that the generated cytology images are quite similar to real breast cytology samples by measuring FID and KID scores.
\keywords{Histopathology  \and Cytology \and CycleGAN \and Unpaired image to image translation.}
\end{abstract}
\section{Introduction}
Automated malignancy identification from breast histopathology images or cytology images is quite challenging. Last few decades many researchers have been trying to develop CAD (Computer Aided Diagnosis) based tools\cite{das2023cervical,mitra2021cytology} for automated diagnosis of malignancy. Due to the lack of expertise in the cytology or histopathology domain, a CAD-based system will be helpful for our society. In recent years, deep learning techniques have given some promising performances in the medical imaging domain. In that case, we need a large amount of annotated datasets, but which are unavailable.  Nowadays, some techniques like data augmentation\cite{dey2020syncgan}, transfer learning\cite{khan2019novel}, and ensemble techniques\cite{dey2023gc} are introduced to overcome these problems. Previously, some traditional data augmentation techniques like random rotation, flipping, cropping, adding noise, etc. were used to boost the classification performances in the medical image analysis domain. Now, various deep learning-based models like variational autoencoder\cite{madan2022synthetic}, Generative Adversarial Neural Network(GAN)\cite{chen2022generative}, Diffusion Model\cite{kebaili2023deep}, etc. are introduced for realistic synthetic data generation, and these synthetic data are combined with a training set to boost the classification performance. These generative models are not only used for data augmentation tasks; it have various tasks like image-to-image translation\cite{abdelmotaal2021pix2pix}, super-resolution\cite{ledig2017photo},  etc. In the task of image-to-image translation, some generative models like pix2pix\cite{isola2017image}, CycleGAN\cite{zhu2017unpaired}, etc. are giving some promising performances. In recent years, transferring from one domain to another has been an important task. In the medical imaging domain, it will be quite helpful. Previously, some works\cite{kearney2020attention,welander2018generative} were proposed where synthetic CT images are generated from MRI images. Also, there are some works in the pathology domain\cite{teramoto2021mutual,runz2021normalization}, where it is used either to transfer from one stain to another stain or to stain normalisation among images. 
In this manuscript, we will make a study where one type of biopsy image is translated to another type of biopsy image. 
Cancer is a deadly disease in today's world. When the patient sees any lumps or tumors in any part of the organ he or she comes to the doctor, and the doctor suggests for biopsy test. There are many biopsy techniques like tissue biopsy, aspiration biopsy, etc. Among different types of biopsy, the most common and least harmful is FNAC(Fine Needle Aspiration Cytology). Among different types of cancers, the most common type for women is breast cancer. According to the survey, it is found that there are many datasets available on breast histopathology domains like BreakHis, BACH, Tiger Challange, etc. But in the cytology domain, only cervical cytology image datasets like HervLev, SipakMed, etc. are publicly available. In the breast cytology domain, there is no publicly available dataset. So, in this study, we have explored image-to-image translation models to create synthetic cytology images from breast histopathology images.

The contributions of the proposed study are:

\begin{enumerate}
    \item Generate realistic synthetic cytology images by translating histopathology images.
\item Generated synthetic cytology images, which have more similar features to the original cytology images.
\end{enumerate}

In the subsequent section, some state-of-the-art methods, related to unpaired image-to-image translation are described. Further, the methodology of the proposed empirical studies is mentioned, and finally, the results are analysed.

\section{Literature Study:}

Previously, many unpaired image-to-image translation models have been explored in the medical image analysis domain. Sometimes generative models are used to make multi-modal data by using a single modality. There are a lot of works reported where computed tomography(CT) images are transformed into magnetic resonance(MR) images(and vice versa) by an unpaired image-to-image translation model. Welander et al.\cite{welander2018generative} proposed a study where they used CycleGan and the UNIT model for unpaired CT to MR translation. Kearney et al.\cite{kearney2020attention} implemented a MR to CT image translation model to reduce the burden of the patients, who required radiation therapy. They have implemented attention attention-aware discriminator network for cycleGAN-based synthetic data generation. It has achieved structural similarity index 0.77 and PSNR 62.35. Along with the radiology domain, there are some works of image-to-image translation in the digital pathology domain. Teramoto et al.\cite{teramoto2021mutual} proposed a stain conversion model by using CycleGAN model.  For the cytology images, the papanicolaou and giemsa stain have many complementary roles, so transferring the stain mutually is an important task in analyzing digital cytology images. Also, there exists another unpaired image translation model to generate synthetic images. Tang et al.\cite{tang2021attentiongan} proposed AttentionGAN model for unpaired image-to-image translation task, also implemented an attention-guided discriminator to consider only the attention regions.

\section{Methodologies}

In this manuscript, we have explored the GAN-based cytology image generation technique from histopathology images. Here, we have introduced two traditional image generative models like CycleGAN\cite{zhu2017unpaired} and Neural Style transfer\cite{gatys2016image}, to translate images from one domain to another domain. 

\subsection{Dataset Descriptions:}
In this study, we have used one breast histopathology image dataset-BreakHis\footnote[1]{$https://web.inf.ufpr.br/vri/databases/breast-cancer-histopathological-database-breakhis/$} and one breast cytology image dataset-JUCYT\footnote[2]{${https://github.com/DVLP-CMATERJU/JUCYT_V1}$}. 

BreakHis is a publicly available breast histopathology image dataset. For this study, we have only used 400x magnification images, of which 588 are benign samples and 1232 are malignant. The images are in resolution of 700 $\times$ 460 pixels. 

JUCYT is a breast cytology image dataset that consists of 169 samples, of which 75 and 94 are Benign and Malignant samples respectively. These samples were collected from Theism Medical Diagnostic Centre, Kolkata, in the presence of professional practitioner. First, the slides by FNAC(Fine Needle Aspiration Cytology) test are collected and viewed under microscope at 40x magnification, and after that, the important regions of interest are captured by the camera.

\begin{figure}[h]
    \centering
    \includegraphics[width=\textwidth]{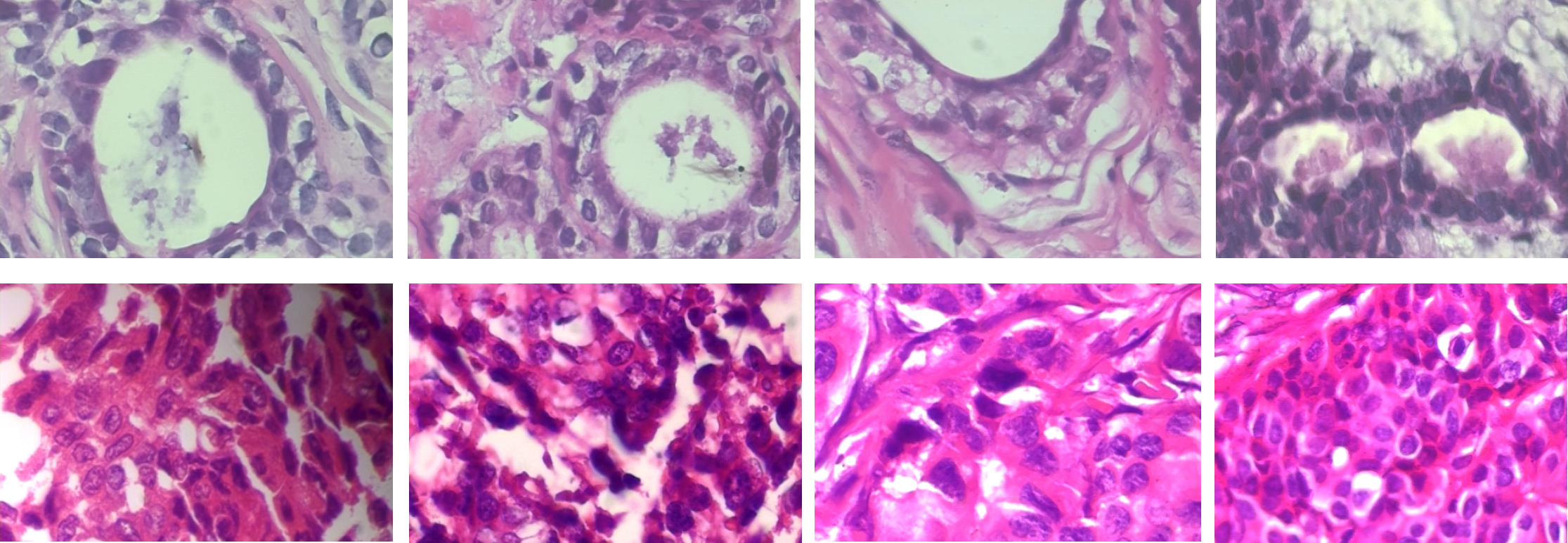}
    \caption{Examples of Real Breast Histopathology samples(from BreakHis Dataset). First row: Benign samples and Second row: Malignant samples.}
    \label{fig:f1}
\end{figure}

\begin{figure}[h]
    \centering
    \includegraphics[width=\textwidth]{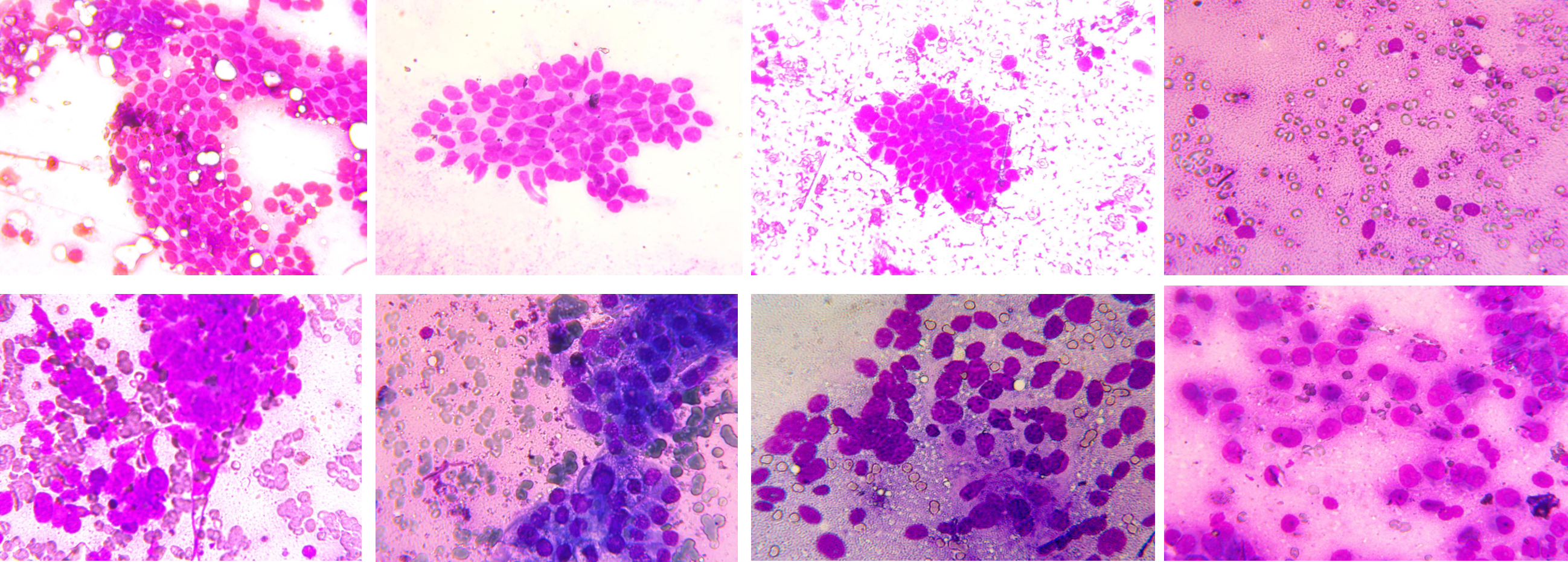}
    \caption{Examples of Real Breast Cytology samples(from JUCYT Dataset). First row: Benign samples and Second row: Malignant samples.}
    \label{fig:f2}
\end{figure}

These two datasets are split into 4:1 ratio for train and test sets. The details data descriptions are mentioned in the following table(Table \ref{tab:t1}). Some real cytology and histopathology samples are mentioned in Fig. \ref{fig:f2} and Fig. \ref{fig:f1} respectively.

\begin{table}[h]
\caption{Data Distributions of Breast Histopathology and Cytology Datasets}
\label{tab:t1}
\resizebox{0.6\columnwidth}{!}{%
\begin{tabular}{|c|cc|cc|}
\hline
\multirow{2}{*}{\textbf{Dataset}} & \multicolumn{2}{c|}{\textbf{Train}} & \multicolumn{2}{c|}{\textbf{Test}} \\ \cline{2-5} 
 & \multicolumn{1}{c|}{\textbf{Benign}} & \textbf{Malignant} & \multicolumn{1}{c|}{\textbf{Benign}} & \textbf{Malignant} \\ \hline
\textbf{BreakHis} & \multicolumn{1}{c|}{473} & 986 & \multicolumn{1}{c|}{115} & 246 \\ \hline
\textbf{JUCYT} & \multicolumn{1}{c|}{55} & 70 & \multicolumn{1}{c|}{20} & 24 \\ \hline
\end{tabular}%
}
\end{table}

From the above figures, it is observed that, in histopathology samples, there are some tissue regions, but in cytology samples, the cellular structures are present. So, from histopathology image to cytology image synthesis is quite a challenging task. So, in the next section, some methodologies of synthetic cytology image generation are explained briefly.

\subsection{Histopathology to Cytology image translation using CycleGAN:}

In the GAN-based image-to-image translation model, we need to pair source and target images, i.e. for the training process target images are needed for corresponding source images. But for the medical imaging domain, specially for the digital pathology domain, target ground truth mask preparation is quite costly. CycleGAN model can work with an unpaired image-to-image translation model to mitigate this issue. 
Let $A$ be the source domain(Histopathology) and $B$ be the target domain(Cytology). The main target is to learn a mapping $G: A \rightarrow B$, where the distribution of data from $G(A)$ will be similar like the distribution of data from $B$. In cycleGAN model, there also another mapping $F$ is defined
where $F: B\rightarrow A$. This inverse mapping is used for generated the synthesis images of source domain. For this study, at testing time, we only translate the images from domain $A$ to domain $B$. Here, two discriminator and two generator models are used for training. Let $D_1$ be a discriminative model which will discriminate between $G(a)$ and $b$ where, $a\in A$ and $b\in B$. For the inverse task, let's define another discriminative model $D_2$, where it's aim to distinguish between $a$ and $F(b)$. Then the adversarial loss for the mapping $G$ is 

$Loss_{G}(G,D_1,A,b) = Min_G Max_{d_1} \mathbb{E}_{b}[\log D_1(b)] + \mathbb{E}_{a}[\log (1-D_1(G(a)))] $. 

The generator $G$ tries to generate fake samples of target domain $G(a)$(i.e. cytology), which will look similar to data from domain $B$, but $D_1$ tries to distinguish between the translated sample $G(a)$ and real $b$.

Another loss, i.e. Cycle Consistency Loss is used on this cycleGAN model, where
$Loss_cyc = \mathbb{E}_{a}[ |F(G(a))-a|] + \mathbb{E}_{b}[ |G(F(b))-b|]$
 
 \textbf{Evaluation of CycleGAN model:}

 At training time, the hyper-parameters like batch size, number of epochs, learning rate are set as 1, 100, 0.0002 respectively. PatchGAN discriminator and ADAM are used as discriminator model and optimizer respectively for training cycleGAN model.  The model is trained on NVIDIA GTX Geforce 970 GB GPU with Intel core-i5 and RAM 16 GB. The model is trained separately for benign and malignant classes. The training time required for each class is approximately 6 hours.

\subsection{Histopathology to Cytology image translation using Neural Style Transfer:}

Neural style transfer is a type of image translation technique, where the synthetic samples are generated by combining the content of source images and the style of target images. In this study, we have used histopathology images as content images and tried to capture the style of cytology images. So, as for the style images, we have randomly chosen the images from the cytology samples.

\section{Result Analysis \& Discussion}

\begin{figure}[h]
    \centering
    \includegraphics[width=\textwidth]{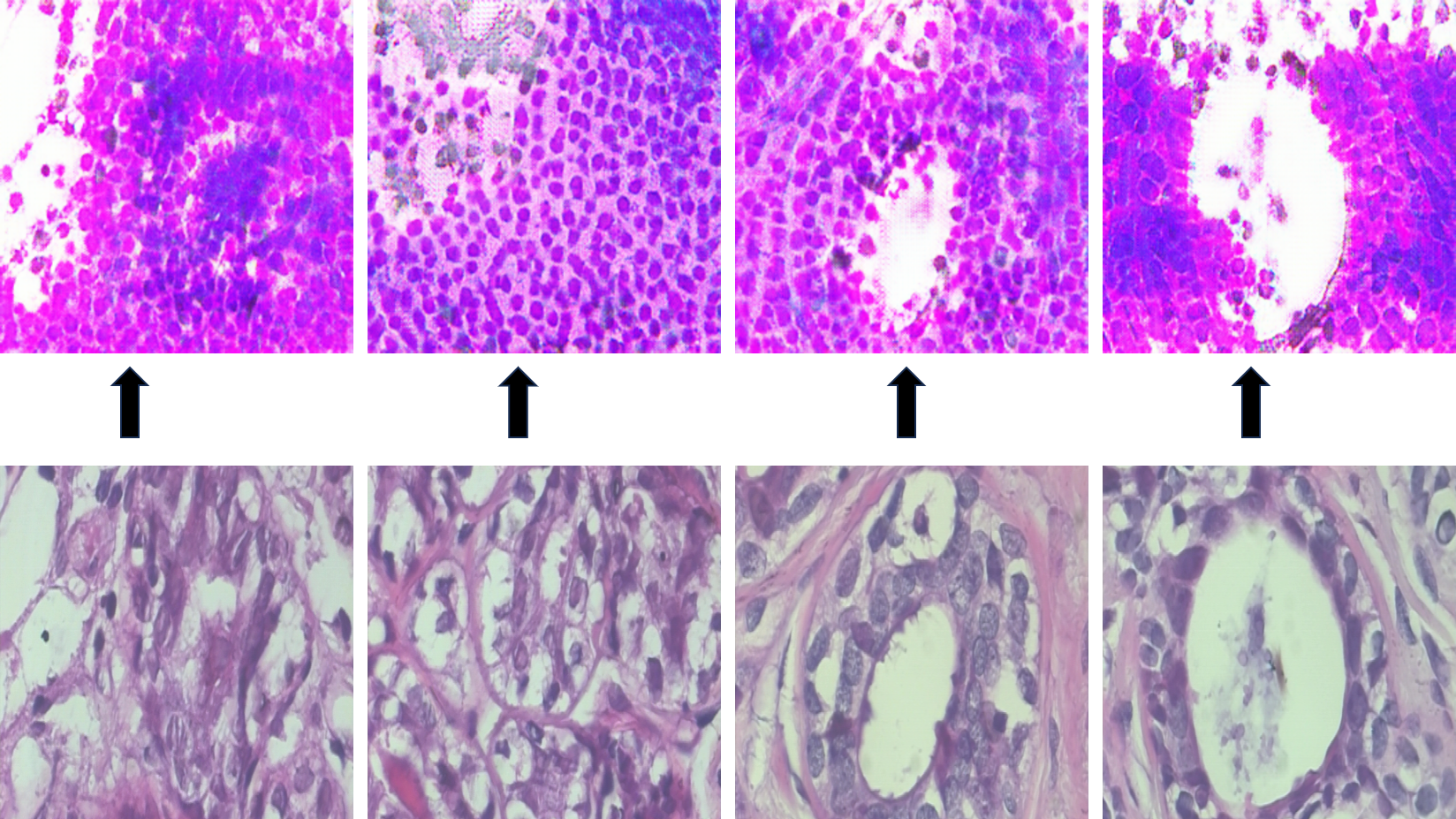}
    \caption{Synthetic Benign Cytology images by CycleGAN model. The second row indicates the histopathology images(Source Domain) and the first row indicates the corresponding synthetic cytology images(Target Domain)}
    \label{fig:f3}
\end{figure}

\begin{figure}[h]
    \centering
    \includegraphics[width=\textwidth]{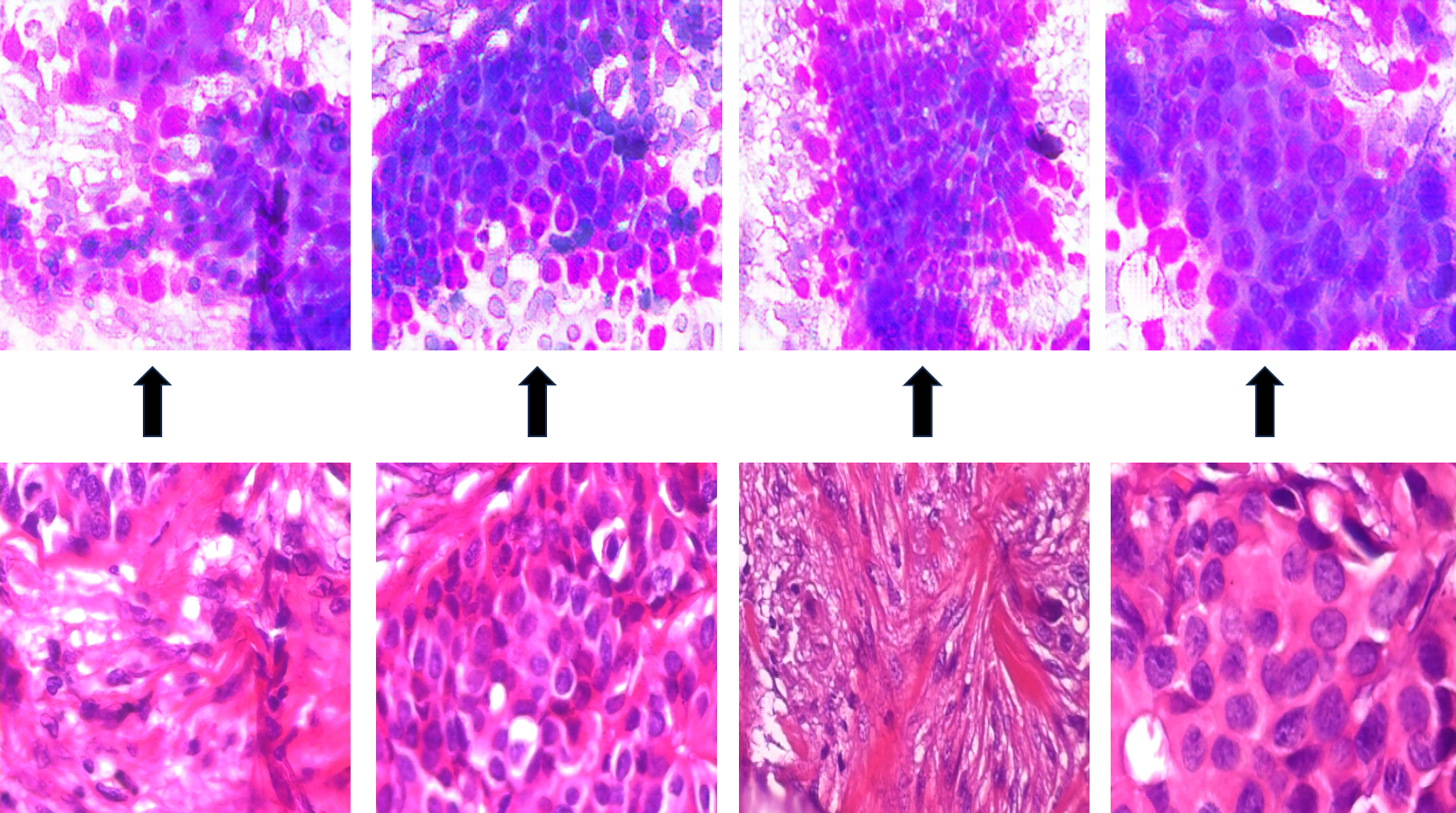}
    \caption{Synthetic Malignant Cytology images by CycleGAN model. The second row indicates the histopathology images(Source Domain) and the first row indicates the corresponding synthetic cytology images(Target Domain)}
    \label{fig:f4}
\end{figure}

\begin{figure}[h]
    \centering
    \includegraphics[width=\textwidth]{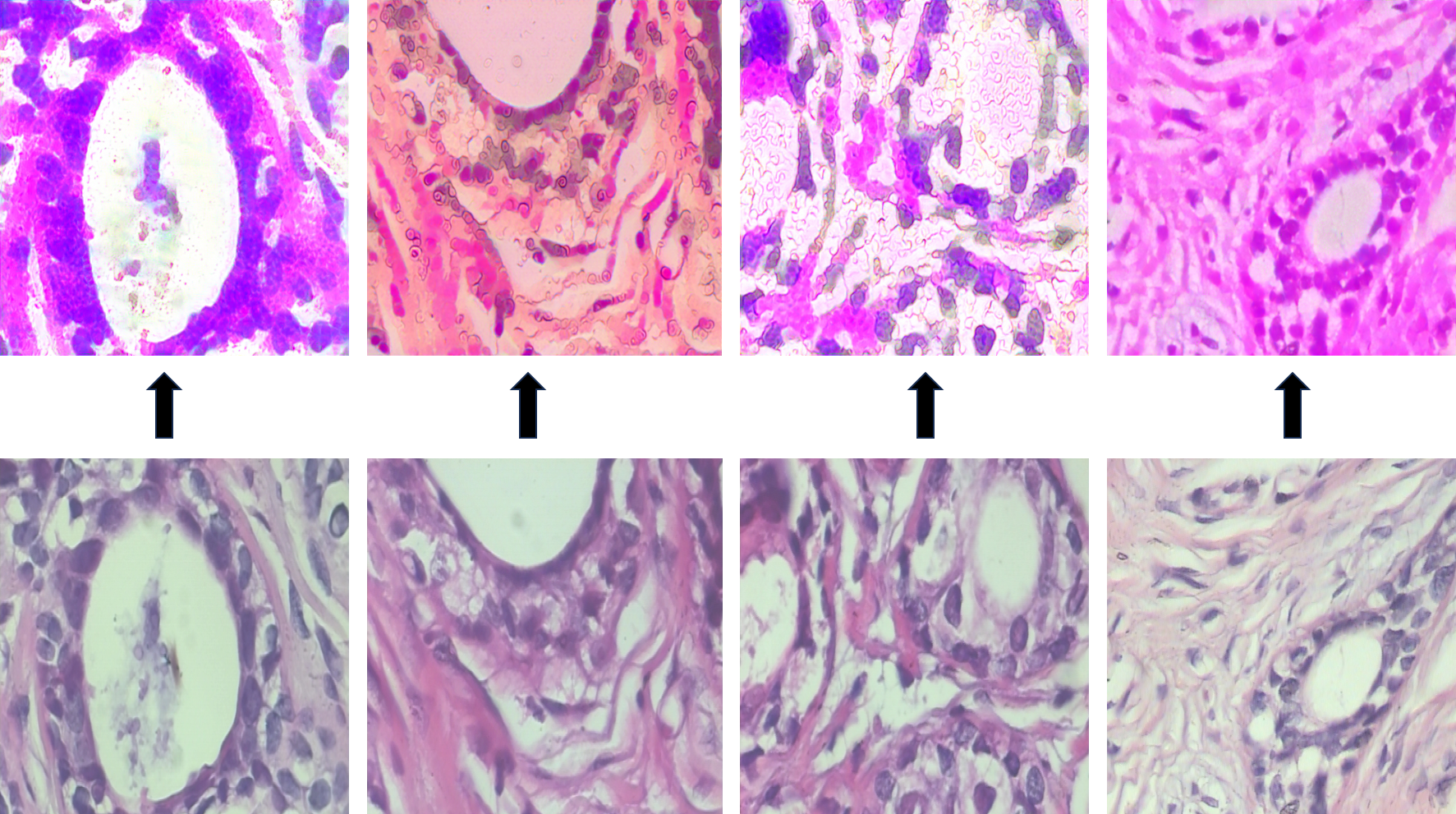}
    \caption{Synthetic Benign Cytology images by Neural Style Transfer model. The second row indicates the histopathology images(Content Image) and the first row indicates the corresponding synthetic cytology images}
    \label{fig:f5}
\end{figure}

\begin{figure}
    \centering
    \includegraphics[width=\textwidth]{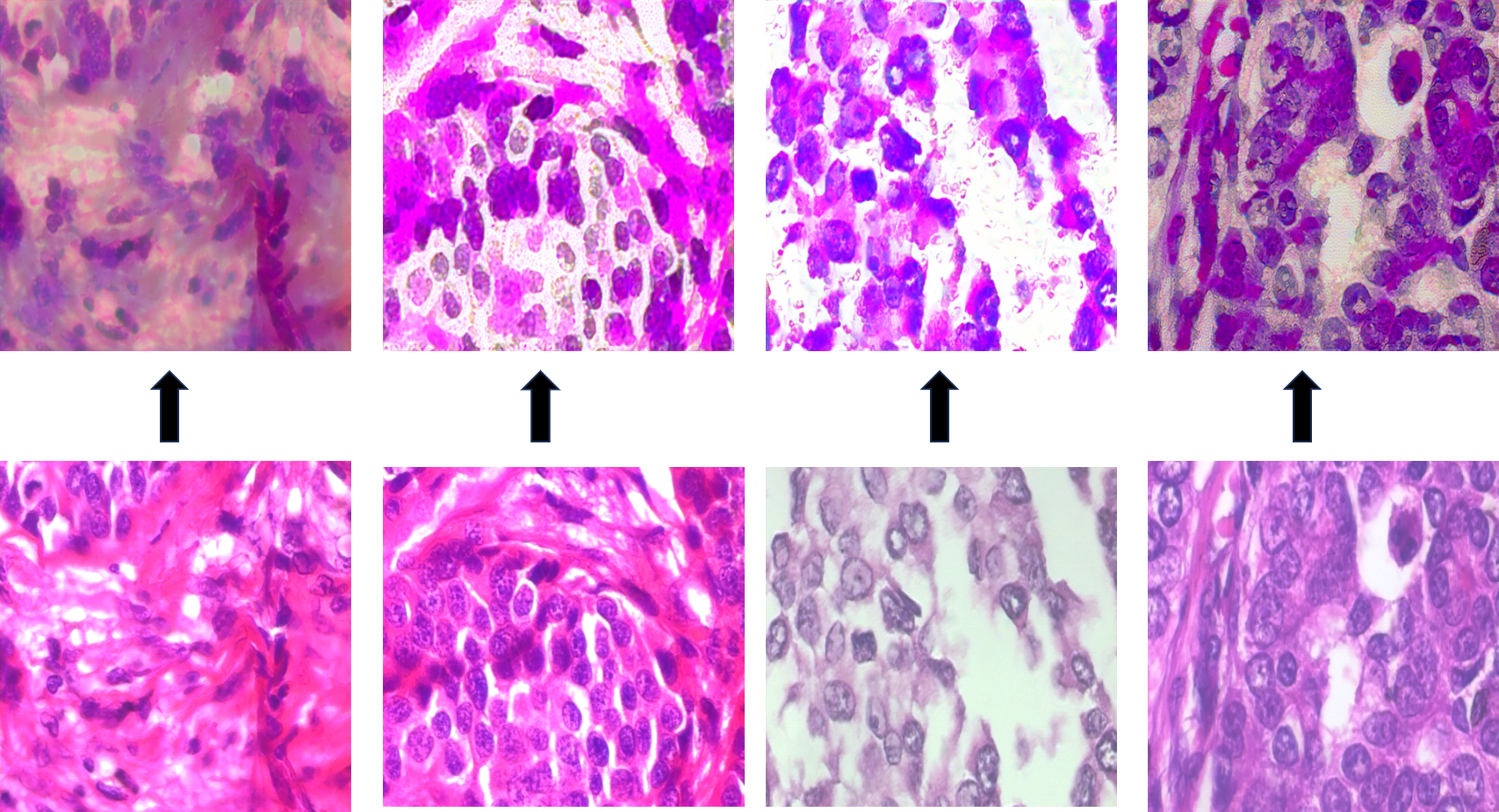}
    \caption{Synthetic Malignant Cytology images by Neural Style Transfer model. The second row indicates the histopathology images(Content Image) and the first row indicates the corresponding synthetic cytology images}
     \label{fig:f6}
\end{figure}

In this study, we have explored image-to-image translation model, to generate synthetic breast cytology images, from the publicly available breast histopathology image datasets. In cytology, images have many cellular objects like nuclei, cytoplasm, red blood cells, etc present. and they are located in proper biological order. Also, for the histopathology images, there are some tissue regions present, like stromal region, mitosis, tumor regions, etc. So, from the perspective of morphological information, they are totally different. In this work, we are trying to translate from one domain to another domain by generative models. On Fig.\ref{fig:f3} and Fig.\ref{fig:f4}, we have shown synthetic Benign and Malignant cytology samples respectively. Also, it is shown that from which histopathology image the synthetic images are transformed. We have made a comparative study concerning the quality between real data and synthetic data distributions. Here we have used two traditional quality matrices FID(Frechet Inception Distance)\cite{yu2021frechet} score and KID(Kernel Inception Distance)\cite{binkowski2018demystifying} score values, to check how much synthetic samples are similar to the original cytology samples. In Table \ref{tab:t2}, we have calculated the FID and KID scores between real cytology and fake cytology samples. Also in Table \ref{tab:t3}, we have calculated scores between real histopathology and synthetic cytology samples. It was found that the FID and KID scores between real cytology and synthetic cytology samples are lower than the scores between real histopathology and synthetic cytology. It is observed that, though the synthetic samples are translated from breast histopathology samples, the feature similarity is much better with real cytology samples than the histopathology samples. But visually, it is observed that the cytology samples capture the semantic structures of objects of histopathology samples. The synthetic samples by neural style transfer model are described in Fig. \ref{fig:f5} and Fig. \ref{fig:f6}. Also, we have measured the quality of fake cytology samples generated by neural style transfer with the original cytology images in Table \ref{tab:t4}. It is observed that the synthetic images generated by CycleGAN are better in quality than the samples generated by neural style transfer. By visual interpretation, we can observe that, in the neural style transfer model, the synthetic samples only capture the style of real cytology images, like background and stain colour, but the main morphological informations like cell size, and cluster of nuclei do not captures, so it is not like an original cytology sample. 

\begin{table}[h]
\caption{Quality metrics between Real Histopathology and Fake Cytology images(generated by CycleGAN model)}
\label{tab:t3}
\resizebox{0.6\columnwidth}{!}{%
\begin{tabular}{|c|c|c|}
\hline
\textbf{Class} & \textbf{FID Score} & \textbf{KID Score} \\ \hline
\textbf{Benign} & 277.589 & 0.0968 $\pm$  0.0419 \\ \hline
\textbf{Malignant} & 222.420 & 0.0582 $\pm$ 0.0423 \\ \hline
\end{tabular}%
}
\end{table}

\begin{table*}[]
\caption{Quality metrics between Real Cytology and Fake Cytology images(generated by CycleGAN model)}
\label{tab:t2}
\resizebox{0.6\columnwidth}{!}{%
\begin{tabular}{|c|c|c|}
\hline
\textbf{Class} & \textbf{FID Score} & \textbf{KID Score} \\ \hline
\textbf{Benign} & 203.904 & 0.0163 $\pm$  0.0295 \\ \hline
\textbf{Malignant} & 143.429 & 0.0018 $\pm$ 0.0154 \\ \hline
\end{tabular}%
}
\end{table*}

\begin{table}[]
\caption{Quality metrics between Real Cytology Samples and Fake Cytology Samples generated by Neural Style Transfer.}
\label{tab:t4}
\resizebox{0.6\columnwidth}{!}{%
\begin{tabular}{|c|c|c|}
\hline
\textbf{Class} & \textbf{FID Score} & \textbf{KID Score} \\ \hline
\textbf{Benign} & 217.58 & 0.1126 $\pm$  0.0324 \\ \hline
\textbf{Malignant} & 199.483 & 0.0763 $\pm$ 0.028 \\ \hline
\end{tabular}%
}
\end{table}

\section{Conclusion}

In this study, we have used two image-to-image translation model to generate breast cytology samples from the real histopathology samples. However at the time of the study, it was observed that the CycleGAN model would be helpful for the unpaired image-to-image translation, but it only generated a finite number of samples. Since it is a transformation from histology to cytology, most of the samples capture the morphological characteristics of cytology samples, but the main problem is that some samples violate the property of benign and malignant samples of cytology images. In the neural style transfer model, the histology samples only capture the styles so, it will not be helpful to generate realistic cytology samples. In future, we will try to explore a transfer learning-based generative model to shift from one domain to another domain.

\section{Acknowledgements}

The authors are thankful to "Theism Medical Diagnosis Centre" for providing the breast cytology slides. This work is supported by SERB, Govt. of India (Ref. no.EEQ/2018/000963).

%
%
%
 \bibliographystyle{splncs04}
 \bibliography{ref}

\end{document}